\documentclass[amsmath,amssymb,aps,pra,twocolumn]{revtex4-2}
\usepackage{txfonts}
\usepackage{bm}
\usepackage{graphicx}

\newcommand{\be}{\begin{eqnarray}}
\newcommand{\ee}{\end{eqnarray}}
\newcommand{\bmat}{\left(\begin{array}}
\newcommand{\emat}{\end{array}\right)}
\newcommand{\no}{\nonumber}

\begin{document}

\title{Bifurcation-based quantum annealing with nested spins}
\author{Kazutaka Takahashi}
\affiliation{Institute of Innovative Research, Tokyo Institute of Technology, Kanagawa 226--8503, Japan}

\date{\today}

\begin{abstract}
We study a bifurcation mechanism of quantum annealing.
Using spins with quantum number $S=1$, 
we construct a simple model to make a bifurcation.
The qutrit can be composed by nesting two qubits.
We numerically solve the Schr\"odinger equation 
to confirm that the bifurcation-based quantum annealing (BQA) works well 
and the ground state can be found efficiently.
The result is compared with that by the standard quantum annealing (QA)
using qubits.
We find that the performance of the BQA
is comparable to the standard QA, or gives better results in some cases.
\end{abstract}

\maketitle

%%%%%%%%%%%%%%%%%%%%%%%%%%%%%%%%%%%%%%%%%%%%%%%%%%%%%%%%%%%%%%%%%%%%%%%%%%%
\section{Introduction}

Quantum annealing (QA) is a heuristic method for solving 
optimization problems~\cite{KN,BBRA}.
It is a kind of adiabatic quantum optimization
algorithms~\cite{FGGS,FGGLLP,AL18} and
is used for a device manufactured 
by D-Wave Systems Inc.~\cite{Jetal,BRIWWLMT}.

In the standard QA, the problem part of the Hamiltonian is represented 
by an Ising-spin model and the quantum fluctuations are induced 
by a transverse-field term.
The corresponding Hamiltonian is familiar in statistical mechanics 
and is used as a standard model for quantum phase transitions~\cite{SIC}.

The transverse field is not the only possible way of
controlling the adiabatic state 
and we can find many other choices in principle.
In fact, it has been recognized that ``nonstoquastic'' effect 
improves the performance~\cite{FGG,BDOT,SN,CFLLS}.
Although it is an interesting problem to find an efficient
driver term from a theoretical point of view,
the implementation of the complicated form of the Hamiltonian
in laboratory is a difficult problem. 

It is also an interesting problem to study
other possible mechanisms utilizing quantum effects.
In this paper, we propose and study 
a bifurcation-based QA (BQA) by using a spin model.
The bifurcation mechanism 
was proposed in a parametrically driven Kerr nonlinear oscillator
as a method of adiabatic quantum optimization~\cite{Goto16-1}.
Goto and his colleagues studied the performance of the
mechanism in Ref.~\cite{Goto16-1} and 
subsequent studies~\cite{Goto16-2,GLN,Goto19,GTD}.
The model is described by bosonic operators and has
continuous degrees of freedom.
It is an interesting problem to find the corresponding mechanism 
in discrete spin models, which is the main aim of this study.

The qubit operations are described by Pauli operators of spin-$1/2$.
Since the operators are too simple to make a bifurcation, 
we consider a higher spin system.
By referring to the standard form of the QA, we construct 
a spin model as a possible realization of the BQA.
We show that the system can be realized in the present technology
and study the performance numerically in the present work.

The organization of this paper is as follows.
In Sec.~\ref{sec:bqa}, we introduce a spin model
realizing a bifurcation and discuss a possible implementation.
In Sec.~\ref{sec:nonint}, we numerically study the bifurcation
mechanism by using a noninteracting Hamiltonian.
The interactions are introduced 
in Sec.~\ref{sec:interaction} and we compare the result
with that from the standard QA.
The last section~\ref{sec:conclusion} is devoted to conclusion.

%%%%%%%%%%%%%%%%%%%%%%%%%%%%%%%%%%%%%%%%%%%%%%%%%%%%%%%%%%%%%%%%%%%%%%%%%%%
\section{Bifurcation-based quantum annealing}
\label{sec:bqa}

%%%%%%%%%%%%%%%%%%%%%%%%%%%%%%%%%%%%%%%%%%%%%%%%%%%%%%%%%%%%%%%%%%%%%%%%%%%
\subsection{Bifurcation mechanism}

The main aim of the QA is to find the ground state of the Hamiltonian 
\be
 H_{\rm p} = -\sum_{\langle i,j\rangle}J_{ij}S_iS_j
 -\sum_{i=1}^N h_iS_i, \label{Hp}
\ee
for a given set of $\{J_{ij}\}$ and $\{h_i\}$.
$\{S_i\}_{i=1,2,\dots,N}$ represents spin variables and 
each spin $S_i$ takes $+1$ or $-1$.
The solution, the ground-state configuration,
is specified by a set of values of $\{S_i\}$.

In the bifurcation mechanism, we start the time evolution from
a symmetric state ``$|0\rangle$'' and 
find degenerate states ``$|\!\pm\! 1\rangle$'' at the end of the evolution.
The degenerate states represent qubit states.
In the standard QA, the initial state is given by a superposition of
final degenerate states: 
$|0\rangle=\left(|\!+\! 1\rangle+|\!-\! 1\rangle\right)/\sqrt{2}$.
To make the bifurcation, we need a bifurcation operator 
that gives the same eigenvalue when it acts on $|\!\pm\! 1\rangle$. 
Since we cannot construct such an operator in qubit systems,
we extend the spin space.

We consider the spin-1 operators 
$\hat{\bm{S}}=(\hat{S}^x,\hat{S}^y,\hat{S}^z)$.
These operators obey the standard commutation relations 
such as $[\hat{S}^x,\hat{S}^y]=i\hat{S}^z$,
and have the quantum number $S=1$ when 
the eigenvalue of $\hat{\bm{S}}^2$ is denoted as $S(S+1)$.
We use the eigenstates of $\hat{S}^z$, $|m\rangle$, as  
\be
 \hat{S}^z|m\rangle = m|m\rangle,
\ee
with $m=+1, 0, -1$.
In this basis, each operator can be represented as
\be
 \hat{S}^z=\bmat{ccc} 1 & 0 & 0 \\ 0 & 0 & 0 \\ 0 & 0 & -1 \emat, \quad
 \hat{S}^x=\frac{1}{\sqrt{2}}\bmat{ccc} 0 & 1 & 0 \\ 1 & 0 & 1 \\ 0 & 1 & 0 \emat.
\ee
Since we do not use $\hat{S}^y$ in the following analysis, 
it is omitted here.
A crucial difference from the Pauli operators is that 
the square of each operator is not proportional to
the identity operator and gives a new kind of operators:
\be
 (\hat{S}^z)^2=\bmat{ccc} 1 & 0 & 0 \\ 0 & 0 & 0 \\ 0 & 0 & 1 \emat, \quad
 (\hat{S}^x)^2=\frac{1}{2}\bmat{ccc} 1 & 0 & 1 \\ 0 & 2 & 0 \\ 1 & 0 & 1 \emat.
 \label{matrices}
\ee

For a single qutrit $i$, we consider the Hamiltonian
\be
 \hat{H}_i(t) = -A(t)\hat{S}_i^x-B(t)(\hat{S}_i^z)^2. \label{Ht}
\ee
We change $B(t)$ slowly from a negative large value 
to a positive large one. 
$A(t)$ is taken to be small but finite values at intermediate times
so that it induces energy-level mixing.
By evolving the system adiabatically with this Hamiltonian, 
we find that the ground state is changed 
from $|0\rangle$ to $|\!\pm\! 1\rangle$.
In the following, we refer to the first term of Eq.~(\ref{Ht})
as driver part and the second term as bifurcation part.

We set the total Hamiltonian for $N$ qutrits as 
\be
 \hat{H}(t) = \sum_{i=1}^N\hat{H}_i(t)
 +\hat{H}_{\rm p}, \label{H}
\ee
where $\hat{H}_{\rm p}$ represents the problem part 
replaced $\{S_i\}$ in Eq.~(\ref{Hp}) with $\{\hat{S}_i^z\}$.
We set $|B(0)|\sim |B(t_{\rm f})|\gg |J_{ij}|\sim |h_i|$ where 
$t_{\rm f}$ represents the annealing time.
Then, each qutrit basically changes from $|0\rangle$ 
to $|\!\pm\! 1\rangle$.
The degeneracy of the final state is lifted by the presence of $\hat{H}_{\rm p}$
and we can solve the optimization problem.

We note that the problem part, $\hat{H}_{\rm p}$, is independent of $t$.
When $B(0)$ is a negative large number and $A(0)$ is negligible, 
the initial state is given by the eigenstate of $\hat{S}_i^z$
with the eigenvalue 0.
The problem part only gives a zero contribution and
does not affect the state even if we keep $\hat{H}_{\rm p}$
from the beginning.
This is one of advantages of the present method.
The time dependence of the Hamiltonian is only on each spin, $\hat{H}_i(t)$,
and we do not need to change the intricate problem part, $\hat{H}_{\rm p}$. 
Then, it is expected that the dynamical property is basically determined
by the driver and bifurcation parts 
and is insensitive to the complexity class of the problem.

%%%%%%%%%%%%%%%%%%%%%%%%%%%%%%%%%%%%%%%%%%%%%%%%%%%%%%%%%%%%%%%%%%%%%%%%%%%
\subsection{Spin coupling by nesting}

One of promising methods realizing the qutrit is 
to use spin nesting.
The sum of two qubits gives 
\be
 \hat{\bm{S}}_i=\frac{1}{2}\left(
 \hat{\bm{\sigma}}_{i1}+\hat{\bm{\sigma}}_{i2}\right),
\ee
where $\hat{\bm{\sigma}}=(\hat{\sigma}^x,\hat{\sigma}^y,\hat{\sigma}^z)$
represents the set of Pauli operators.
According to the principle of quantum mechanics, 
$\hat{\bm{S}}_i$ represents operators
with the quantum number $S=0$ or 1.
When we set the initial state as an eigenstate with $S=1$, 
the Hamiltonian does not change $S$ and 
the state of the system is described by qutrit, three of four states.

\begin{center}
\begin{figure}[t]
\begin{center}
\includegraphics[width=0.6\columnwidth]{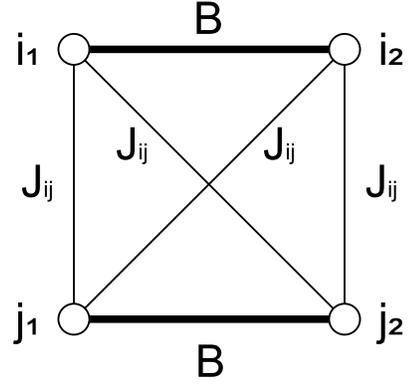}
\end{center}
\caption{The connectivity graph of two logical qutrits by four physical qubits.
Logical qutrit $i$ is made from physical qubits $i1$ and $i2$,
and qutrit $j$ from qubits $j1$ and $j2$.
Two physical qubits within a single qutrit interact with each other 
and the interaction is controlled by $B(t)$.
The interaction between two qutrits in the problem Hamiltonian, $J_{ij}$, 
is represented by four bonds. 
We also need additional single-qubit operations as represented by 
$A(t)$ (for $\sum_i\hat{\sigma}_i^x$) and $h_i$
(for $\hat{\sigma}_{i_1}^z+\hat{\sigma}_{i_2}^z$).
}
\label{fig1}
\end{figure}
\end{center}

The connectivity of two qutrits is specified in Fig.~\ref{fig1}.
A single logical qutrit is made from two physical qubits.
The driving represented by $B$ is achieved 
by operating the interaction between 
physical qubits within a single qutrit.
The interaction between two qutrits, $J_{ij}$, 
is represented by four bonds.

It is interesting to find that the present method
is equivalent to nesting for 
an error-proofing procedure~\cite{MNAL,MNVAL,Matsuura}.
It is expected that the nested qubit can be robust against noise 
due to the ferromagnetic coupling between the physical qubits.
In our choice of the Hamiltonian in Eq.~(\ref{H}), 
$\hat{S}_i^x$ changes the states $|\!\pm\! 1\rangle_i$ to $|0\rangle_i$
and there is no direct transition between
$|\!+\!1\rangle_i$ and $|\!-\!1\rangle_i$.
Since the energy level of the state $|0\rangle_i$ becomes
large at large $t$, quantum fluctuations
represented by virtual transitions to different levels are suppressed, 
which might be related to an error-proofing property.
We note that the behavior can be changed by
introducing additional driver terms.
For example, $(\hat{S}_i^x)^2$ gives a direct coupling between
$|\!+\!1\rangle_i$ and $|\!-\!1\rangle_i$.
We see from Eq.~(\ref{matrices}) that 
$(\hat{S}^x)^2$ is equivalent to $\hat{\sigma}^x$
if the Hilbert space is effectively restricted to $m=\pm 1$.
%Since we want to keep the robust structure of the driver part,
%we do not introduce additional terms in the following calculations.

We note that the initial state for each qutrit is given by $|0\rangle_i$
with $\hat{S}^z_i|0\rangle_i=0$.
It can be written by qubit states as 
\be
 |0\rangle_i=\frac{1}{\sqrt{2}}\left(
 |\!+\!1/2\rangle_{i1}\otimes|\!-\!1/2\rangle_{i2}
 +|\!-\!1/2\rangle_{i1}\otimes|\!+\!1/2\rangle_{i2}
\right).
\ee
$|\!\pm\! 1/2\rangle$ represent two qubit states.
Although this is an entangled state and 
cannot be obtained by a single qubit operation,
the manipulation is only for two qubits 
and can be obtained, e.g., by the standard QA procedure.
We know various ways of controlling systems with a small number of spins
and it is expected that the state can be prepared efficiently.

%%%%%%%%%%%%%%
%%%%%%%%%%%%%%

%%%%%%%%%%%%%%%%%%%%%%%%%%%%%%%%%%%%%%%%%%%%%%%%%%%%%%%%%%%%%%%%%%%%%%%%%%%
\section{Noninteracting systems}
\label{sec:nonint}

We study the performance of the BQA 
by solving the Schr\"odinger equation numerically.
In this section, we treat noninteracting systems 
to confirm that the bifurcation mechanism works efficiently.
Each qutrit can be treated independently and the mechanism
can be studied by the single qutrit Hamiltonian 
\be
 \hat{H}(t)=-A(t)\hat{S}^x-B(t)(\hat{S}^z)^2. \label{H1}
\ee

\begin{center}
\begin{figure}[t]
\begin{center}
\includegraphics[width=0.8\columnwidth]{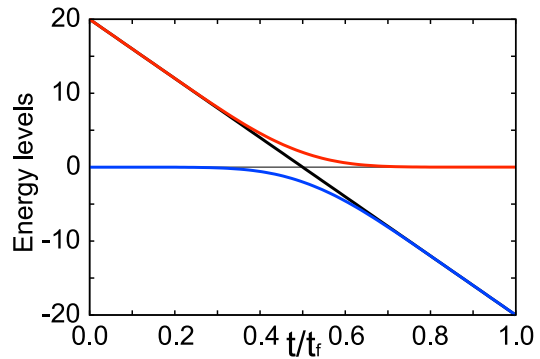}
\end{center}
\caption{
The instantaneous energy levels of the Hamiltonian in Eq.~(\ref{H1}).
We use Eqs.~(\ref{B}) and (\ref{Ag}), and take $B_0/A_0=20$.
The energy levels are plotted in unit of $A_0$.
}
\label{fig2}
\end{figure}
\end{center}

We use the linear protocol for $B(t)$:
\be
 B(t)=B_0 \left(2\frac{t}{t_{\rm f}}-1\right), \label{B}
\ee
where $B_0$ is a positive constant much larger than $A(t)$.
Since our method is based on adiabaticity,
we take $t_{\rm f}$ to be a large value.
As we mentioned in the previous section, 
$A(t)$ takes small but finite values at intermediate times.
We use the Gaussian protocol  
\be
 A(t)= A_0 \exp\left[-\frac{1}{2\sigma^2}\left(2\frac{t}{t_{\rm f}}-1\right)^2\right], \label{Ag}
\ee
with $\sigma^2=0.1$.
The instantaneous energy levels of the Hamiltonian in Eq.~(\ref{H1}) 
are plotted in Fig.~\ref{fig2}.
The energy gap between the ground state 
and the excited state at $t=0$ is very large.
After passing through avoided-crossing region 
around $t/t_{\rm f}= 0.5$, 
the system has the ground state with two-fold degeneracy. 

There is no guiding principle on the choice of $A(t)$.
In the following, we also examine the case when $A(t)$ takes a constant value 
because the time-independent protocol is practically convenient.
Although $A(0)$ must be zero so that the state becomes 
an eigenstate of $\hat{S}^z$ at $t=0$, 
it is enough provided $|B(0)|\gg |A(0)|$ is satisfied.

\begin{center}
\begin{figure}[t]
\begin{center}
\includegraphics[width=0.8\columnwidth]{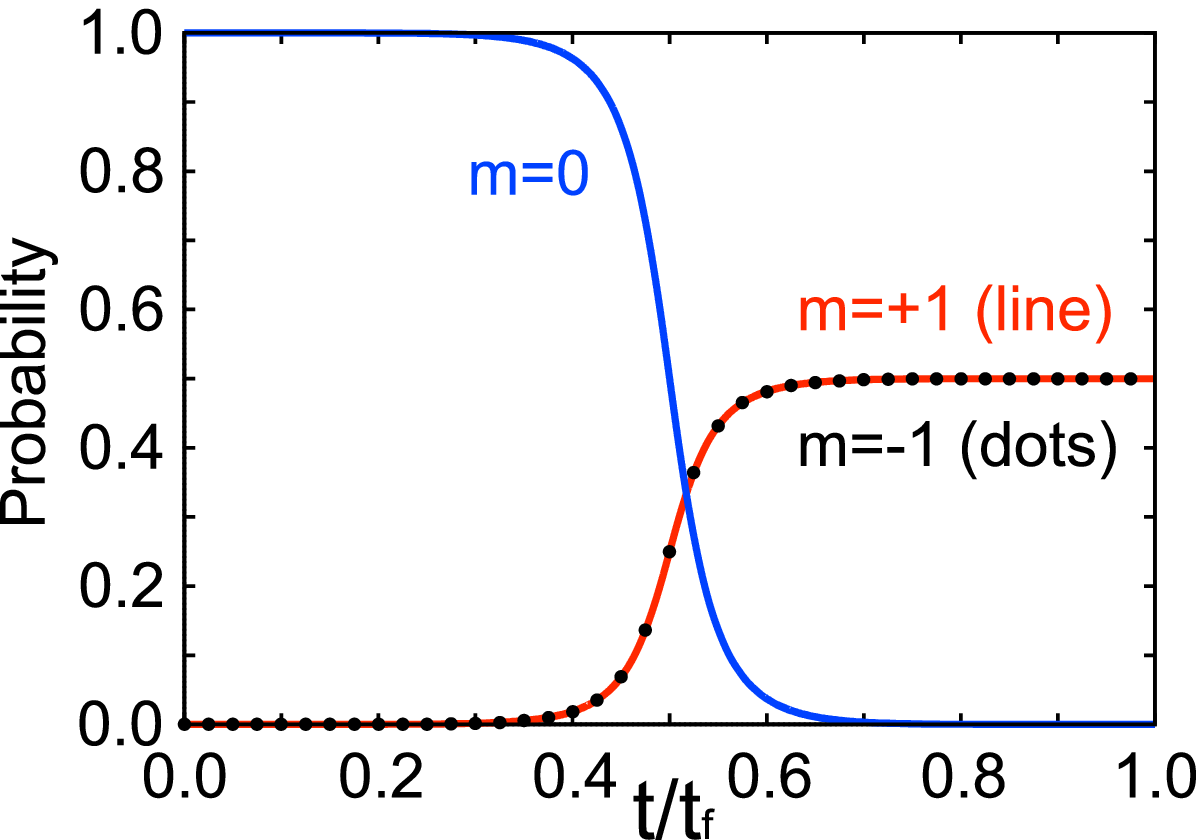}  
\includegraphics[width=0.8\columnwidth]{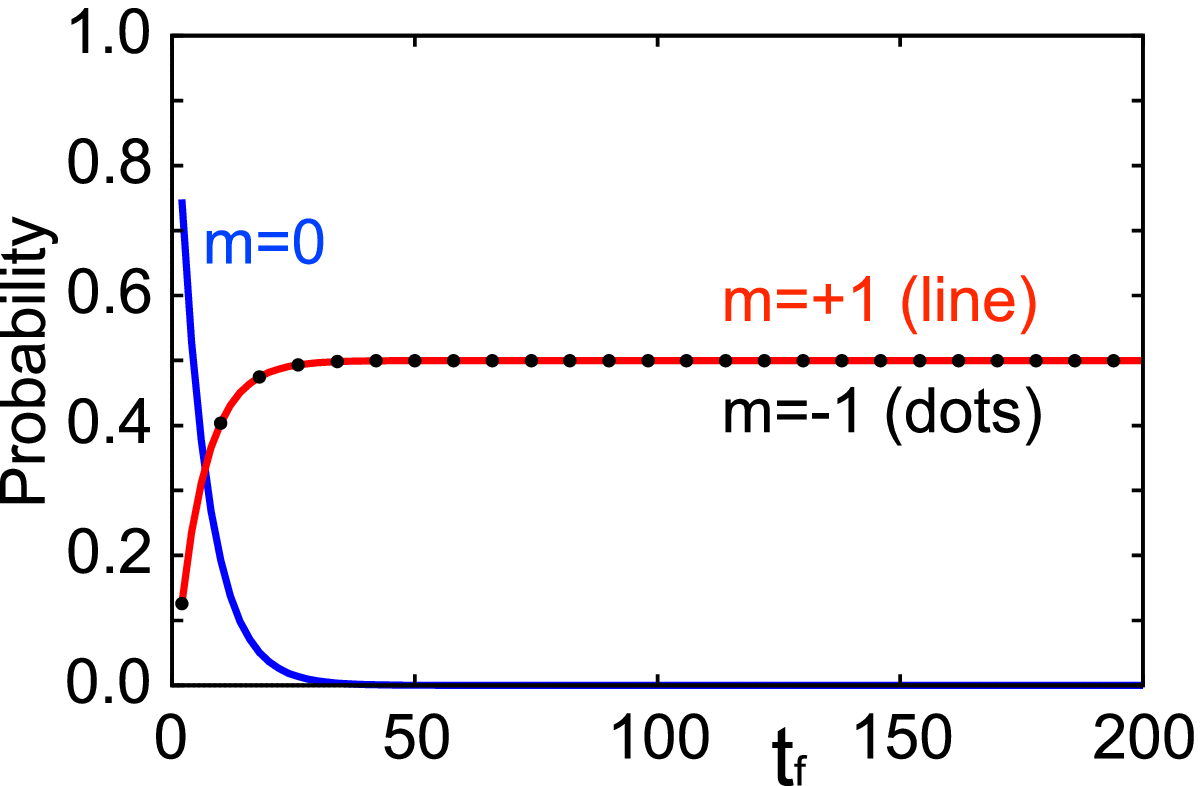}
\end{center}
\caption{
The solution of the Schr\"odinger equation with
the Hamiltonian in Eq.~(\ref{H1}). 
We use the protocol in Eqs.~(\ref{B}) and (\ref{Ag}), 
and take $B_0/A_0=20$.
Top: The probability distributions of the time-evolved state at each $t$.
We take $A_0t_{\rm f}=100$.
Bottom: The annealing-time dependence of the final state.
Here, $t_{\rm f}$ is plotted in unit of $A_0$.
}
\label{fig3}
\end{figure}
\end{center}

We numerically solve the Schr\"odinger equation with 
the Hamiltonian in Eq.~(\ref{H1}) 
to obtain the time-evolved state $|\psi(t)\rangle$.
We first use the Gaussian protocol in Eq.~(\ref{Ag}).
In Fig.~\ref{fig3}, we plot the time dependence of
probabilities $|\langle m|\psi(t)\rangle|^2$ with $m=+1,0,-1$ 
for a given $t_{\rm f}$, 
and the annealing-time dependence of 
$|\langle m|\psi(t_{\rm f})\rangle|^2$.
We see that the bifurcation mechanism works very well 
if the annealing time is not considerably small.
The final state is given by 
$\left(|\!+\!1\rangle+|\!-\!1\rangle\right)/\sqrt{2}$ 
and has components of $m=\pm 1$ with equal probability.

\begin{center}
\begin{figure}[t]
\begin{center}
\includegraphics[width=0.8\columnwidth]{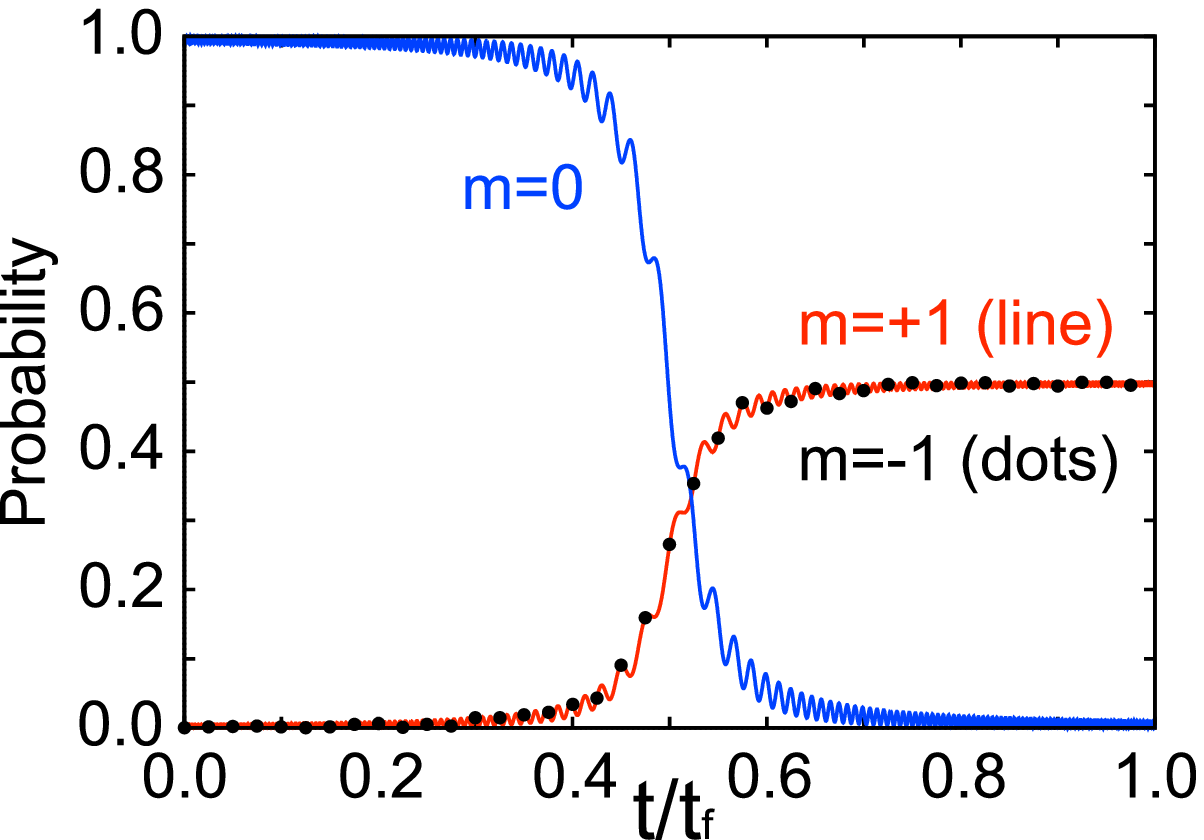}
\includegraphics[width=0.8\columnwidth]{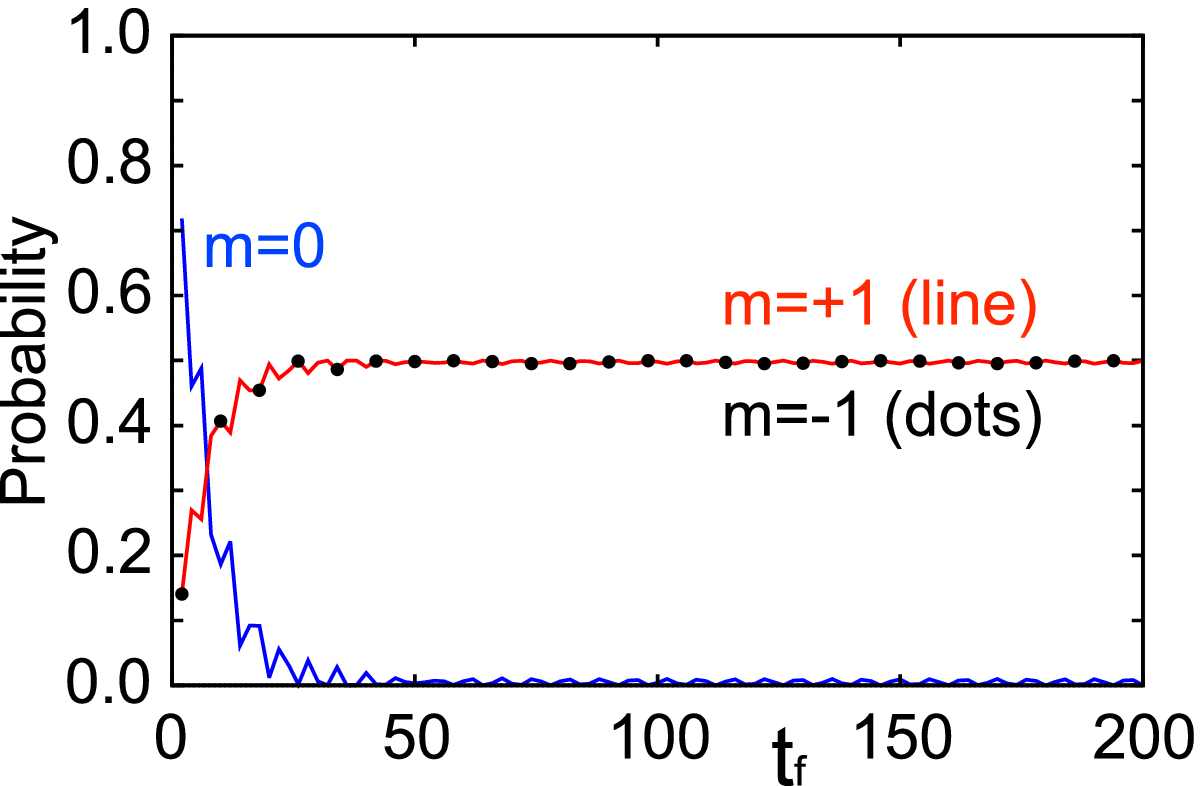}
\end{center}
\caption{
The solution of the Schr\"odinger equation with the Hamiltonian
in Eq.~(\ref{H1}).
We take $A(t)=A_0={\rm const}.$.
The other parameters are the same as those in Fig.~\ref{fig3}.
}
\label{fig4}
\end{figure}
\end{center}

We consider the case where $A(t)$ is constant: $A(t)=A_0$.  
we plot the result in Fig.~\ref{fig4}.
Although we see small oscillations, 
the performance is almost the same as that in Fig.~\ref{fig3}.
We also examined several other cases and found similar results.
This implies robustness of the bifurcation mechanism.

Next, we incorporate the noninteracting part of $\hat{H}_{\rm p}$. 
We put $J_{ij}=0$, which means that 
we still have a noninteracting system and
the single qutrit Hamiltonian is given by 
\be
 \hat{H}(t)=-A(t)\hat{S}^x-B(t)(\hat{S}^z)^2-h\hat{S}^z, \label{Hh}
\ee
where $h$ represents the magnetic field.
The final result is determined by the sign of $h$.

\begin{center}
\begin{figure}[t]
\begin{center}
\includegraphics[width=0.8\columnwidth]{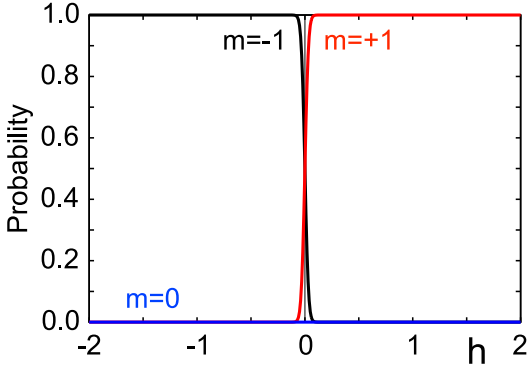}
\end{center}
\caption{
The performance with the Hamiltonian in Eq.~(\ref{Hh}).
We use Eqs.~(\ref{B}) and (\ref{Ag}) with $B_0/A_0=20$, and $A_0t_f=200$.
Each curve represents the probability of the component $m$ at $t=t_{\rm f}$.
Here, $h$ is plotted in unit of $A_0$.
}
\label{fig5}
\end{figure}
\end{center}

The result is plotted in Fig.~\ref{fig5}.
We see that that the proper state, $m=+1$ or $-1$, 
is selected as a function of $h$, 
if $|h|$ is not too small.

%%%%%%%%%%%%%%
%%%%%%%%%%%%%%

%%%%%%%%%%%%%%
%%%%%%%%%%%%%%
%%%%%%%%%%%%%%
%%%%%%%%%%%%%%
%%%%%%%%%%%%%%
%%%%%%%%%%%%%%

%%%%%%%%%%%%%%%%%%%%%%%%%%%%%%%%%%%%%%%%%%%%%%%%%%%%%%%%%%%%%%%%%%%%%%%%%%%
\section{Interacting systems}
\label{sec:interaction}

Having confirmed that the bifurcation mechanism works well 
for a single qutrit, we study multi qutrit systems with interactions.

%%%%%%%%%%%%%%%%%%%%%%%%%%%%%%%%%%%%%%%%%%%%%%%%%%%%%%%%%%%%%%%%%%%%%%%%%%%
\subsection{Ferromagnetic interactions}

We first consider ferromagnetic interactions for 
a one-dimensional arrangement of spins with periodic boundary condition.
Each spin interacts with the neighboring spins
and we set $J_{1,2}=J_{2,3}=\cdots=J_{N,1}=J>0$ and $h_i/J=0.1$.
Here, we introduce a finite $h_i$ to avoid degenerate ground states. 
The effect of degeneracy is discussed in the next subsection.

To assess the performance of the BQA, 
we compare the result with 
that of the standard QA:
\be
 \hat{H}(t) &=& \left(1-\frac{t}{t_{\rm f}}\right)
 \left(-\Gamma\sum_{i=1}^N\hat{\sigma}_i^x\right) \no\\
 && +\frac{t}{t_{\rm f}}\left(
 -\sum_{\langle i,j\rangle}J_{ij}\hat{\sigma}^z_i\hat{\sigma}^z_j
 -\sum_{i=1}^N h_i\hat{\sigma}_i^z\right). \label{Hqa}
\ee
Each element is represented by qubit and 
the standard linear protocol is used to control the system.

\begin{center}
\begin{figure}[t]
\begin{center}
\includegraphics[width=0.8\columnwidth]{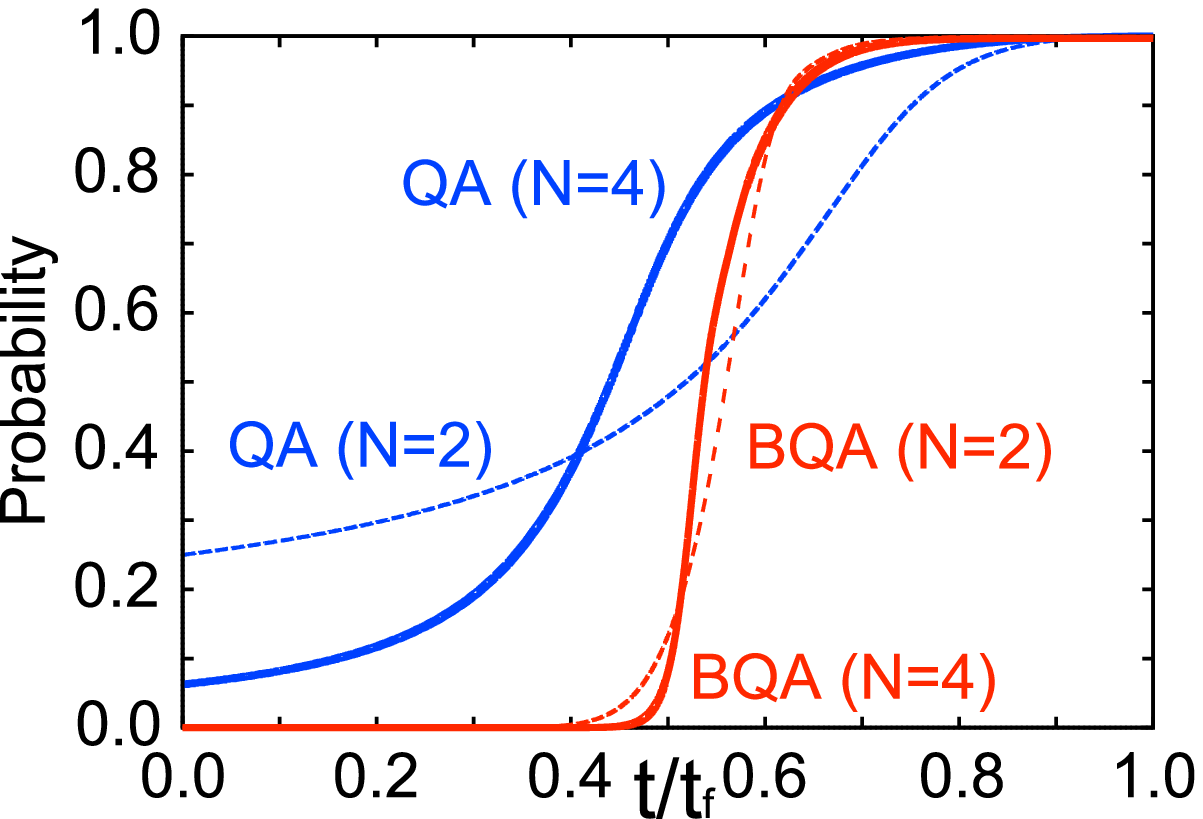}
\includegraphics[width=0.8\columnwidth]{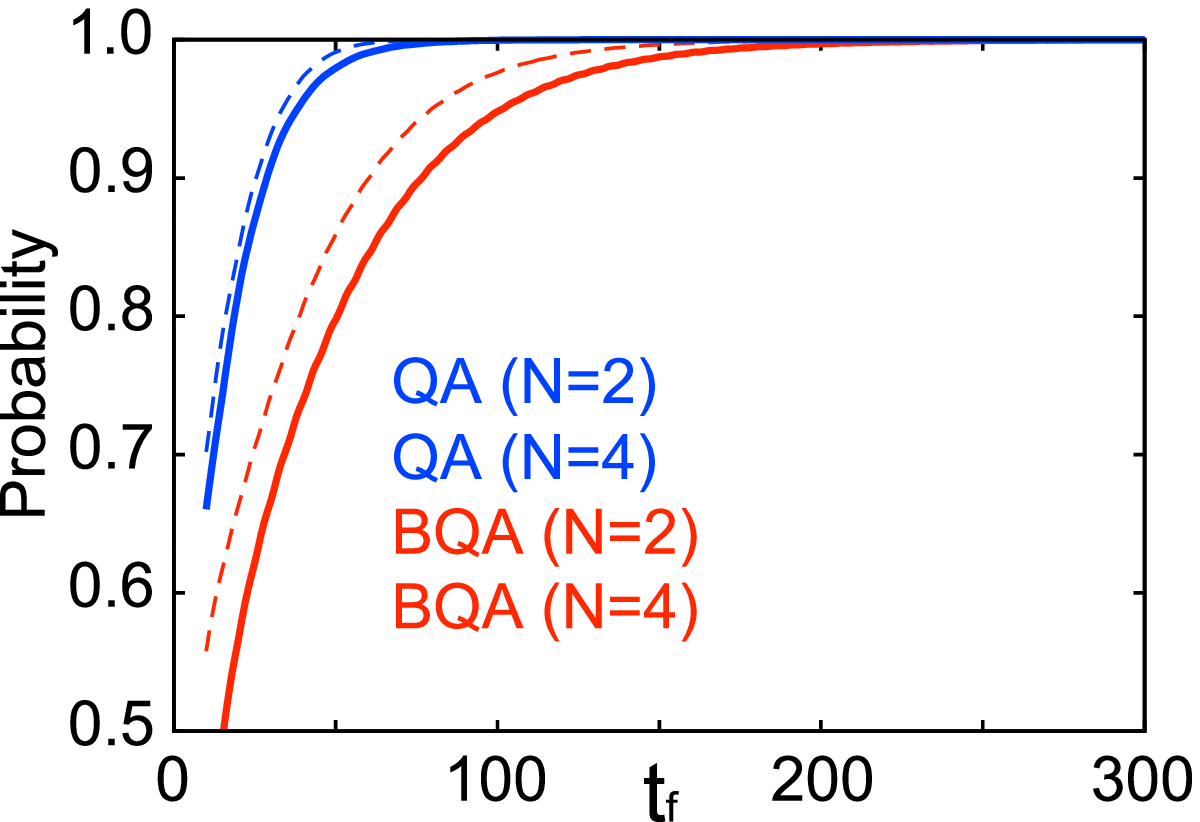}
\end{center}
\caption{
The performance with the ferromagnetic problem Hamiltonian
specified in the text.
For the BQA, we use the protocol in Eqs.~(\ref{B}) and (\ref{Ag}) 
with $B_0/J=20$ and $A_0/J=2$.
For the QA, we use Eq.~(\ref{Hqa}) with $\Gamma/J=1$.
Top: The time dependence of the ground-state probability.
We take $Jt_{\rm f}=200$.
Bottom: The annealing-time dependence.
Here, $t_{\rm f}$ is plotted in unit of $J$.
}
\label{fig6}
\end{figure}
\end{center}

The numerical result is plotted in Fig.~\ref{fig6}.
We see that, in our choice of the parameters,
the computation works very well.
In contrast to the standard QA,
the initial state with $m=0$ is changed 
to the final one abruptly after $t$ exceeds $t_{\rm f}/2$.
When $t$ is much smaller than $t_{\rm f}/2$, the bifurcation part 
is the dominant contribution and 
the state remains the zero state.
After passing through the region where the driver part is dominant, 
the state is changed to the ground state of $\hat{H}_{\rm p}$.

Comparison between the QA and the BQA 
in the bottom panel of Fig.~\ref{fig6}  shows 
that a large annealing time is required 
to obtain the ideal result in the case of the BQA.
In the present implementation of the QA~\cite{Jetal,BRIWWLMT}, 
the scale of the Hamiltonian is of the order of GHz, and the annealing time 
is of the order of $\mu$s.
This corresponds to $Jt_{\rm f}\sim 1000$ in our unit,
which is large enough to find the ideal result.

%%%%%%%%%%%%%%
%%%%%%%%%%%%%%

%%%%%%%%%%%%%%%%%%%%%%%%%%%%%%%%%%%%%%%%%%%%%%%%%%%%%%%%%%%%%%%%%%%%%%%%%%%
\subsection{(Un-)Fair sampling property}

In the previous example, we used a problem Hamiltonian with 
no ground-state degeneracy.
The standard QA is known to give a biased sampling
among the degenerate ground states~\cite{MNK, KMOKT} 
and we study this property in the BQA.

\begin{center}
\begin{figure}[t]
\begin{center}
\includegraphics[width=0.8\columnwidth]{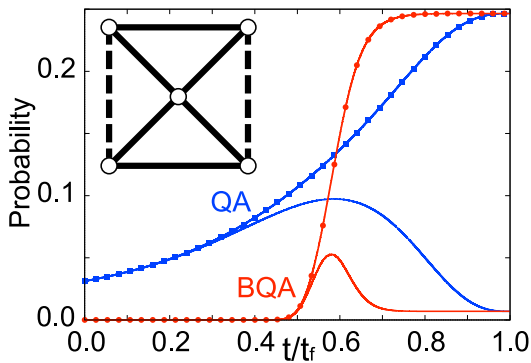}
\end{center}
\caption{
(Un-)Fair sampling properties of the QA and BQA.
We use a five spin model with six degenerate ground states.
The connectivity is specified in the inset where 
solid lines represent ferromagnetic interaction ($J_{ij}=J>0$) and 
dashed lines antiferromagnetic interaction ($-J<0$).
We plot three of six states.
Two of them are plotted by solid lines
and the other is plotted by dotted line.
The blue lines are for QA and the red for BQA.
We take $Jt_{\rm f}=300$.
}
\label{fig7}
\end{figure}
\end{center}

We use a five spin model used in Ref.~\cite{MNK} 
which is denoted in the inset of Fig.~\ref{fig7}.
This system has six ground states.
Half of them are due to spin-flip symmetry and 
we plot three levels in Fig.~\ref{fig7}.
We see that the result of the BQA is very similar to that of the QA.
Two of three levels are equally sampled and 
the other single level is suppressed.
We checked that this property is unchanged
when we use several different protocols.

As discussed in the original study~\cite{MNK},
we can improve the result by introducing
additional driver terms to the Hamiltonian.
Since the present model has a larger Hilbert space, 
we have many choices to improve the result, in principle.
It is an interesting problem, but is beyond the scope of 
the present study.

%%%%%%%%%%%%%%
%%%%%%%%%%%%%%

%%%%%%%%%%%%%%%%%%%%%%%%%%%%%%%%%%%%%%%%%%%%%%%%%%%%%%%%%%%%%%%%%%%%%%%%%%%
\subsection{Random interactions}

We study random systems where $J_{ij}$ and $h_i$ are chosen randomly.
We treat a fully-connected model with $J_{ij}=r_{ij}/N$ ($i\ne j$), and 
$r_{ij}$ and $h_i$ are sampled from uniform distribution $[-J,J]$.

\begin{center}
\begin{figure}[t]
\begin{center}
\includegraphics[width=0.8\columnwidth]{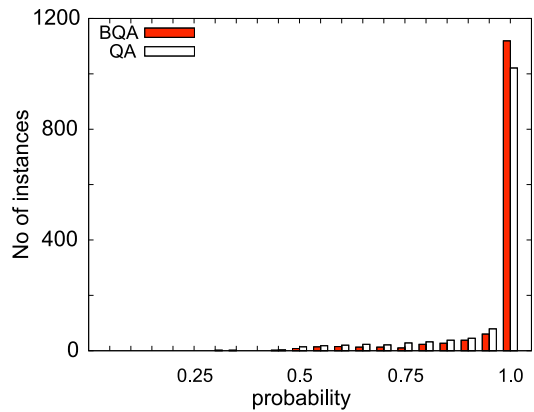}
\end{center}
\caption{
Histograms for the success probability for random Hamiltonians with $N=4$.
We compare the results of the BQA and QA.
We take $Jt_{\rm f}=300$ and the number of samples is 1600.
The bin width of the histogram is 0.05.
}
\label{fig8}
\end{figure}
\end{center}

We show the result in Fig.~\ref{fig8}.
We see that the BQA outperforms the QA,
though we cannot find a drastic change.
We checked in the result of the BQA 
that the obtained state does not include the zero state $|0\rangle_i$,
which means that the bifurcation works well.

\begin{center}
\begin{figure}[t]
\begin{center}
\includegraphics[width=0.8\columnwidth]{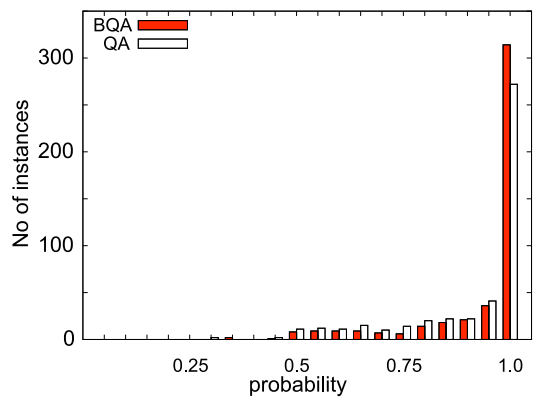}
\end{center}
\caption{
The result for the case 
where the ground-state configuration is nontrivial.
The calculation conditions are the same as those in Fig.~\ref{fig8}.
The number of samples is 454.
}
\label{fig9}
\end{figure}
\end{center}

To see that the method works even if 
the solution of the problem is nontrivial, 
we plot in Fig.~\ref{fig9} the result for samples in which 
the ground-state configurations $\{S_i\}_{i=1,\dots,N}$ are not equal to 
$\{{\rm sign}(h_i)\}_{i=1,\dots,N}$. 
We still find that the BQA gives a better result than the QA.

%%%%%%%%%%%%%%
%%%%%%%%%%%%%%
%%%%%%%%%%%%%%
%%%%%%%%%%%%%%

%%%%%%%%%%%%%%
\begin{center}
\begin{figure}[t]
\begin{center}
\includegraphics[width=0.8\columnwidth]{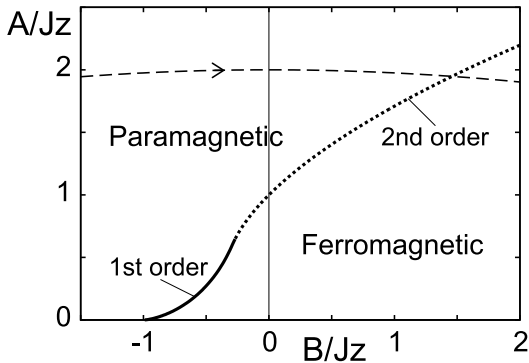}
\end{center}
\caption{
The phase diagram of the ferromagnetic model
in the mean-field approximation.
The paramagnetic phase ($m_{\rm s}=0$) and
ferromagnetic phase ($m_{\rm s}>0$) are
separated by first-order and second-order phase-transition lines.
The dashed line with arrow represents the protocol $(A(t),B(t))$ 
used in this study.
}
\label{fig10}
\end{figure}
\end{center}
%%%%%%%%%%%%%%
%%%%%%%%%%%%%%
\begin{center}
\begin{figure}[t]
\begin{center}
\includegraphics[width=0.8\columnwidth]{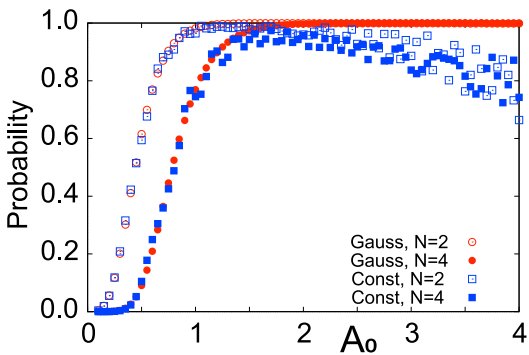}
\end{center}
\caption{
$A_0$ dependence of the result.
We study the ferromagnetic model treated in Fig.~\ref{fig6}.
``Gauss'' represents the protocol $A(t)$ in Eq.~(\ref{Ag}) and
``Const'' represents $A(t)=A_0$.
}
\label{fig11}
\end{figure}
\end{center}
%%%%%%%%%%%%%%

%%%%%%%%%%%%%%%%%%%%%%%%%%%%%%%%%%%%%%%%%%%%%%%%%%%%%%%%%%%%%%%%%%%%%%%%%%%
\subsection{First-order phase transition for large systems}

It is hard to obtain numerical results 
for large values of $N$ in the present method.
Instead, we study statistical properties at thermodynamic limit $N\to\infty$
by using the mean-field approximation.

The statistical model of the present type of the Hamiltonian 
has been discussed in various works 
as a model to describe $\lambda$ transition 
in mixtures of ${\rm He}^3$ and ${\rm He}^4$~\cite{Blume,Capel,BEG,GS}.
In the mean-field approximation for ferromagnetic systems 
without magnetic field, 
the system is described by effective Hamiltonian 
\be
 \hat{H}_{\rm eff}(m_{\rm s})=-A\hat{S}^x-B(\hat{S}^z)^2-Jzm_{\rm s}\hat{S}^z,
\ee
where $z$ represents the coordination number, 
the number of couplings of a single spin to the other spins,
and $m_{\rm s}$ is the magnetization determined selfconsistently.
The selfconsistent equation is written as 
\be
 m_{\rm s} =\langle\psi_{\rm GS}(m_{\rm s})|\hat{S}^z|\psi_{\rm GS}(m_{\rm s})\rangle,
\ee
where $|\psi_{\rm GS}(m_{\rm s})\rangle$ is
the ground state of $\hat{H}_{\rm eff}(m_{\rm s})$.

The selfconsistent equation always has the paramagnetic solution $m_{\rm s}=0$.
The ferromagnetic solutions with $m_{\rm s}>0$ are obtained
in a certain range of parameters as we show in Fig.~\ref{fig10}.
Those two phases are separated by a phase transition.
It is of second order when the order parameter $m_{\rm s}$ changes continuously
and of first order when $m_{\rm s}$ changes discontinuously.
The first-order phase transitions occur
when $|A|$ is small and $B$ is negative.
At the first-order transition, 
the zero state $m=0$ is changed discontinuously to the qubit states.
We note that the paramagnetic phase with $m_{\rm s}=0$
does not distinguish between the zero state $m=0$ and
the Ising paramagnetic state, mixtures of $m=\pm 1$.
The zero state is dominant when $B$ is negative and
the Ising paramagnetic state is dominant when $B$ is positive.

Since the first-order transition is between 
the zero state and the qubit states, 
this property is mainly determined by 
competing effects between the driver part and the bifurcation part
and is insensitive on the details of the problem part.
In fact, we can also find a similar behavior 
when we treat random systems~\cite{GS}.
We still find a first-order transition at small $|A|$ and negative $B$ 
with the ferromagnetic phase replaced by the spin-glass phase.

It is known that the QA fails when the system goes across
the first-order phase boundary~\cite{JKKMP}.
To avoid the first-order transition in the BQA,
$A$ must be taken to be a large value.
We study a ferromagnetic model to see how the result is dependent 
on the choice of $A_0$. 
The result is plotted in Fig.~\ref{fig11}.
The computation fails when $A_0$ is small as we expect from 
the phase diagram in Fig.~\ref{fig10}.
The statistical mechanical analysis shows that
the failure at small $A_0$ 
is restricted to a finite range of the parameter
even if we consider large $N$.

%%%%%%%%%%%%%%%%%%%%%%%%%%%%%%%%%%%%%%%%%%%%%%%%%%%%%%%%%%%%%%%%%%%%%%%%%%%
\section{Conclusion}
\label{sec:conclusion}

We have discussed the bifurcation mechanism by using a spin model.
The model can be constructed from the standard qubit system by nesting.
We found in our numerical calculation 
that the performance of the BQA is better than that of the QA.
Although we did not find a drastic change,
the result can be further improved by optimizing protocols, driver part,
and some other parameters.

Compared with the standard QA, our method has several remarkable properties.
First, the problem part of the Hamiltonian
is independent of time and is convenient for implementations.
We can only control the driver and bifurcation parts
which are common to any process.
Since the dynamics is mainly determined by those parts,
we can study optimizations of the protocol by using
the single qutrit Hamiltonian.

Second, our Hamiltonian forbids direct transition
between qubit states $m\pm 1$.
Their states are only interchanged by way of the zero state.
It is considered to give an error-proofing property.

Third, the model uses an extended Hilbert space and
we can in principle introduce different types of operators
to enhance the performance.
We must be careful when we introduce a new operator
since it can affect the second property we mentioned above.

Admittedly, the present study is limited to small spin systems and
it is difficult to draw firm conclusions on the performance of the BQA.
However, we stress that studying a different mechanism of adiabatic
quantum optimization algorithms is an important problem 
to obtain a better understanding of quantum computations.
We expect that the mechanism discussed in this paper
brings a new direction of research.

%%%%%%%%%%%%%%%%%%%%%%%%%%%%%%%%%%%%%%%%%%%%%%%%%%%%%%%%%%%%%%%%%%%%%%%%%%%%
%\section*{Acknowledgments}
%This work was supported by JSPS KAKENHI Grant 
%Number JP26400385 (K.T.). 
%The author is grateful to Takuya Hatomura for useful comments.

%%%%%%%%%%%%%%%%%%%%%%%%%%%%%%%%%%%%%%%%%%%%%%%%%%%%%%%%%%%%%%%%%%%%%%%%%%%%
%%%%%%%%%%%%%%%%%%%%%%%%%%%%%%%%%%%%%%%%%%%%%%%%%%%%%%%%%%%%%%%%%%%%%%%%%%%%


\section*{References}
\begin{thebibliography}{99}


\bibitem{KN}
T. Kadowaki and H. Nishimori, 
Quantum annealing in the transverse Ising model, 
Phys. Rev. E {\bf 58}, 5355 (1998).

\bibitem{BBRA}
J.~Brooke, D.~Bitko, T.~F.~Rosenbaum, and G.~Aeppli, 
Quantum annealing of a disordered magnet, 
Science {\bf 284}, 779 (1999).

\bibitem{FGGS}
E.~Farhi, J.~Goldstone, S.~Gutmann, and M.~Sipser, 
Quantum computation by adiabatic evolution, 
arXiv: quant-ph/0001106 (2000).

\bibitem{FGGLLP}
E.~Farhi, J.~Goldstone, S.~Gutmann, J.~Lapan, A.~Lundgren, and D.~Preda, 
A quantum adiabatic evolution algorithm applied to random instances of an NP-complete problem, 
Science {\bf 292}, 472 (2001).

\bibitem{AL18}
T. Albash and D. A. Lidar,
Adiabatic quantum computation,
Rev. Mod. Phys. {\bf 90}, 015002 (2018).

\bibitem{Jetal}
M. W. Johnson, et al.,
Quantum annealing with manufactured spins, 
Nature {\bf 473}, 194 (2011).

\bibitem{BRIWWLMT}
S.~Boixo, T.~F.~R\o nnow, S.~V.~Isakov, Z.~Wang, D.~Wecker, D.~A.~Lidar, 
J.~M.~Martinis, and M.~Troyer, 
Evidence for quantum annealing with more than one hundred qubits,
Nat. Phys. {\bf 10}, 218 (2014).

\bibitem{SIC}
S. Suzuki, J. Inoue, and B. K. Chakrabarti, 
Quantum Ising phases and transitions in transverse Ising models, 2nd ed. 
(Springer, 2013).

\bibitem{FGG}
E. Farhi, J. Goldstone, and S. Gutmann,
Quantum adiabatic evolution algorithms with different paths, 
arXiv:quant-ph/0208135 (2002).

\bibitem{BDOT}
S.~Bravyi, D.~P.~DiVincenzo, R.~Oliveira, and B.~M.~Terhal, 
The complexity of stoquastic local Hamiltonian problems, 
Quantum Inf. Comput. {\bf 8}, 361 (2008).

\bibitem{SN}
Y. Seki and H. Nishimori, 
Quantum annealing with antiferromagnetic fluctuations, 
Phys. Rev. E {\bf 85}, 051112 (2012).

\bibitem{CFLLS}
E.~Crosson, E.~Farhi, C.~Y.-Y.~Lin, H.-H.~Lin, and P.~Shor, 
Different strategies for optimization using the quantum adiabatic algorithm, 
arXiv:1401.7320 (2014).

\bibitem{Goto16-1}
H. Goto, 
Bifurcation-based adiabatic quantum computation with a nonlinear oscillator network, 
Sci. Rep. {\bf 6}, 21686 (2016).
%10.1038/srep21686

\bibitem{Goto16-2}
H. Goto, 
Universal quantum computation with a nonlinear oscillator network, 
Phys. Rev. A {\bf 93}, 050301(R) (2016).

\bibitem{GLN}
H. Goto, Z. Lin, and Y. Nakamura, 
Boltzmann sampling from the Ising model using quantum heating of coupled nonlinear oscillators,
Sci. Rep. {\bf 8}, 7154 (2018).

\bibitem{Goto19}
H. Goto,
Quantum computation based on quantum adiabatic bifurcations of Kerr-nonlinear parametric oscillators, 
J. Phys. Soc. Jpn. {\bf 88}, 061015 (2019).

\bibitem{GTD}
H. Goto, K. Tatsumura, and A. R. Dixon, 
Combinatorial optimization by simulating adiabatic bifurcations in nonlinear Hamiltonian systems, 
Sci. Adv. {\bf 5}, eaav2372 (2019).

\bibitem{MNAL}
S.~Matsuura, H.~Nishimori, T.~Albash, and D.~A.~Lidar, 
Mean Field Analysis of Quantum Annealing Correction, 
Phys. Rev. Lett. {\bf 116}, 220501 (2016).

\bibitem{MNVAL}
S.~Matsuura, H.~Nishimori, W.~Vinci, T.~Albash, and D.~A.~Lidar, 
Quantum-annealing correction at finite temperature: Ferromagnetic p-spin models,
Phys. Rev. A {\bf 95}, 022308 (2017).

\bibitem{Matsuura}
S. Matsuura, 
Mean field quantum annealing correction,
J. Phys. Soc. Jpn. {\bf 88}, 061006 (2019).

\bibitem{MNK}
Y.~Matsuda, H.~Nishimori, and H.~G.~Katzgraber,
Ground-state statistics from annealing algorithms: quantum versus classical 
approaches, 
New J. Phys. {\bf 11}, 073021 (2009).

\bibitem{KMOKT}
M.~S.~K\"onz, G.~Mazzola, A.~J.~Ochoa, H.~G.~Katzgraber, and M.~Troyer, 
Uncertain fate of fair sampling in quantum annealing, 
Phys. Rev. A {\bf 100}, 030303(R) (2019).

\bibitem{Blume}
M. Blume,
Theory of the first-order magnetic phase change in UO2,
Phys. Rev. {\bf 141}, 517 (1966).

\bibitem{Capel}
H. W. Capel,
On the possibility of first-order phase transitions in Ising systems
of triplet ions with zero-field splitting,
Physica {\bf 32}, 966 (1966).

\bibitem{BEG}
M. Blume, V. J. Emery, and R. B. Griffiths,
Ising model for the lambda transition and phase separation in 
He3-He4 mixtures,
Phys. Rev. A {\bf 4}, 1071 (1971).

\bibitem{GS}
S. K. Ghatak and D. Sherrington, 
Crystal field effects in a general S Ising spin glass,
J. Phys. C: Solid State Phys. {\bf 10}, 3149 (1977).

\bibitem{JKKMP}
T.~J\"org, F.~Krzakala, J.~Kurchan, A.~C.~Maggs, and J.~Pujos,
Energy gaps in quantum first-order mean-field-like transitions: 
The problems that quantum annealing cannot solve,
Europhys. Lett. {\bf 89}, 40004 (2010).

\end{thebibliography}
\end{document}